\documentclass{article}

\usepackage{microtype}
\usepackage{graphicx}
\usepackage{subcaption}
\usepackage{booktabs}
\usepackage{hyperref}
\usepackage{afterpage}

\usepackage[preprint]{icml2026}

\makeatletter
\renewcommand{\Notice@String}{Preprint.}
\makeatother

\usepackage{amsmath}
\usepackage{amssymb}

\graphicspath{{./}}

\icmltitlerunning{Feature Visualization Recovers Cortical Selectivity from TRIBE v2}

\begin{document}

\twocolumn[
\icmltitle{Feature Visualization Recovers Known\\ Cortical Selectivity from TRIBE v2}

\begin{icmlauthorlist}
\icmlauthor{Stuart Bladon}{duke}
\icmlauthor{Brinnae Bent}{duke}
\end{icmlauthorlist}

\icmlaffiliation{duke}{Duke University, Durham, NC, USA}
\icmlcorrespondingauthor{Stuart Bladon}{stuartbladon@icloud.com}

\icmlkeywords{brain encoders, feature visualization, mechanistic interpretability, fMRI, TRIBE, V-JEPA}

\vskip 0.3in
]

\printAffiliationsAndNotice{}

\begin{abstract}
Brain encoder models predict cortical fMRI responses from the internal activations of pretrained vision and language networks, and are typically evaluated by held-out prediction accuracy. This is a useful signal for training but a poor one for interpretation: it tells us an encoder fits the data without telling us whether it has internalized the functional organization of the brain. We propose feature visualization --- gradient ascent on the encoder's predicted activation for a target region of interest (ROI) --- as a complementary interpretability technique, and apply it to TRIBE v2 composed with V-JEPA 2 (ViT-G, 40 layers), holding both frozen and synthesizing still images for seven regions spanning the ventral and dorsal visual hierarchies. Under identical hyperparameters, the probe recovers a visible progression of increasing spatial scale and feature complexity across V1$\to$V4, matching the ventral-stream hierarchy. It also produces three distinctive downstream regimes: radial ``frozen-motion'' streaks for the middle temporal area (MT) despite static-only optimization, face-like features for the fusiform face area (FFA), and consistent rectilinear line patterns for the parahippocampal place area (PPA). Optimized FFA stimuli drive the predicted region ${\sim}4\times$ as much as a natural face photograph, consistent with feature visualization producing adversarial super-stimuli rather than canonical exemplars. The probe is simple, differentiable, and applicable to any brain encoder with a differentiable backbone, allowing for qualitative evaluation of brain encoders.
\end{abstract}

\section{Introduction}

Brain encoders --- neural networks trained to predict cortical fMRI responses from the internal activations of pretrained vision or language models \citep{yamins2016goaldriven,schrimpf2020brainscore} --- have become a standard tool for studying how artificial representations align with neural ones. They are evaluated almost entirely by held-out prediction accuracy. This is the correct signal for fitting, but it is functionally incomplete. Two encoders that match in held-out $R^2$ can still make systematically different predictions about the input-side preimage of any given cortical region; prediction accuracy on a held-out video corpus does not, by itself, tell us \emph{why} an encoder's predictions are right.

Synthesizing an input that maximally drives a chosen unit, channel, or region by gradient ascent in pixel or Fourier space \citep{erhan2009visualizing,simonyan2013deep,olah2017featurevis} is a useful feature visualization technique to fill that gap. Given a brain encoder $f(\cdot)$ that maps image $x$ to predicted cortical activation $\hat{y}$, gradient ascent on the mean predicted activation within a target region of interest yields a stimulus that, according to the encoder, should maximally drive that region. If the resulting image looks like a face when we target the fusiform face area (FFA), or like oriented edges when we target primary visual cortex (V1), the encoder has recovered a piece of functional anatomy that a neuroscientist would recognize. If it does not, the encoder's held-out accuracy is consistent with a surface-level fit that has not internalized the underlying representations.

We propose feature visualization as an interpretability probe for brain encoders, and apply it to TRIBE v2 (composed with the V-JEPA 2 ViT-G visual backbone it reads from) for seven cortical regions spanning the visual hierarchy: primary visual cortex (V1), secondary visual cortex (V2), third visual area (V3), fourth visual area (V4), middle temporal area (MT), fusiform face area (FFA), and parahippocampal place area (PPA). The probe recovers a visible progression of increasing spatial scale and feature complexity across V1$\to$V4, plus three distinctive downstream regimes that match the canonical selectivity of MT and FFA directly and produce a consistent (but not category-identifiable) signature for PPA. We also describe the training recipe utilized and quantify the gap between optimized stimuli and natural face photographs.

A brain encoder that has recovered functional organization is a much more useful scientific instrument than one that has merely fit the training fMRI time series: it can be queried for what each region ``looks for,'' compared against the neuroscience literature region by region, and used to design stimuli for downstream experiments. The probe we describe is simple to run, requires no training data beyond what the encoder was already fit on, and applies to any encoder with a differentiable backbone, making it a cheap, default qualitative test to run alongside prediction accuracy on every new encoder.

\subsection{Background: what to expect from each region}
The seven regions we target have the following canonical selectivities, summarized here for reference reading the rest of the paper. V1 (primary visual cortex) is tuned to small oriented edges, analogous to Gabor filters \citep{hubel1962receptive}. V2 and V3 progressively enlarge receptive fields and respond to contours, textures, and illusory edges \citep{hegde2000selectivity}. V4 encodes mid-level visual features such as curvature, shape fragments, and color patches \citep{pasupathy2002population}. MT (middle temporal area) is motion-selective: its neurons tune to direction, speed, and optic flow, and typically require moving stimuli \citep{maunsell1983functional,born2005structure}. FFA (fusiform face area) responds preferentially to human faces over other object categories \citep{kanwisher1997fusiform}. PPA (parahippocampal place area) responds to scenes, buildings, landscapes, and spatial layouts \citep{epstein1998cortical}. A successful feature visualization should produce stimuli recognizable by these properties: oriented edges for V1, implied motion for MT, face-like imagery for FFA, and scene or layout cues for PPA.

\section{Related Work}

\paragraph{Brain encoder models.} Predicting fMRI responses from the internal activations of pretrained vision or language networks is a well-established paradigm \citep{yamins2016goaldriven,schrimpf2020brainscore}. TRIBE \citep{tribev1} and its successor TRIBE v2 \citep{tribev2} scale this paradigm to multimodal pretraining on large video foundation models, and are evaluated, like most prior work, by held-out prediction accuracy. The main complement to prediction accuracy in this literature is representational similarity analysis between model and brain. Both ask whether the encoder's representations resemble neural ones at the dataset level, leaving open the input-side question of what each region prefers.

\paragraph{Feature visualization.} Synthesizing inputs that maximally drive chosen units by gradient ascent is a standard interpretability technique for vision networks \citep{erhan2009visualizing,simonyan2013deep,olah2017featurevis}. Common components include Fourier-domain parameterization, decorrelated colour channels, transform-averaging regularization, and low-frequency curricula \citep{olah2017featurevis}. These recipes were developed for unit- or channel-level objectives in image classifiers; their behaviour when applied to whole-brain fMRI encoders --- whose targets are pooled across thousands of cortical vertices and whose backbones are video, not image, models --- is largely unstudied.

\paragraph{Interpretability of brain encoders.} Prior work has used representational similarity analysis, linear probing, and saliency maps to probe what brain encoders have learned. A closer line of work synthesizes images that maximize encoder ROI predictions \citep{walker2019inception,bashivan2019neural,ponce2019evolving,ratanmurty2021computational}, typically using GAN or diffusion priors to keep results on the natural-image manifold. While priors make resulting stimuli easier to read, they also confound regional selectivity with prior guidance: a face-shaped output may reflect the GAN's face manifold rather than the encoder's face tuning. A prior-free probe across the full visual hierarchy of a modern multimodal video encoder is, to our knowledge, missing from the literature.

\section{Method}

\subsection{Models}

\paragraph{V-JEPA 2 ViT-G} \citep[\texttt{facebook/vjepa2-vitg-fpc64-256}, fp16;][]{vjepa2}. A 40-layer vision transformer pretrained on video with a joint-embedding predictive objective \citep{lecun2022jepa}. Hidden size 1408; input 64 frames at $256\times256$; tubelet size 2 yields 32 temporal tokens, and patch size 16 yields 256 spatial tokens per temporal position.

\paragraph{TRIBE v2} \citep[\texttt{facebook/tribev2};][]{tribev2}. A brain encoder that takes per-layer features from V-JEPA 2 (plus optional text/audio features, zero-filled here) and predicts activation at ${\sim}20{,}000$ fsaverage5 cortical vertices at 100 output timesteps. We feed features from two V-JEPA 2 layers (indices 20 and 39, per TRIBE's training config) and treat all modalities but video as zero.

Both models are frozen; only the pixel/Fourier parameters of the image are learned. All seven targets are spatial rather than motion-selective, so we optimize a single still frame and tile it to 64 identical frames at inference. Because V-JEPA 2's tubelet size is 2, only two tubelet-paired frames need to be forwarded, yielding a $32\times$ compute reduction per step at no loss in feature fidelity for static stimuli.

\subsection{ROI definition}
Cortical ROIs are read off the HCP-MMP1 \citep{glasser2016multimodal} parcellation on fsaverage5. The HCP-MMP1 parcels for V1, V2, V3, V4, and MT each map directly to a single eponymous parcel (with 523, 383, 242, 180, and 38 vertices respectively); FFA is taken as the \texttt{FFC} parcel (104 vertices); and PPA is the union of \texttt{PHA1}, \texttt{PHA2}, and \texttt{PHA3} (194 vertices).

\subsection{Image parameterization}
The image is parameterized as a real-valued 2-D Fourier spectrum $S \in \mathbb{R}^{1 \times 3 \times H \times (W/2+1) \times 2}$ on decorrelated colour channels. We apply the inverse real fast Fourier transform (IRFFT) and a sigmoid to map the spectrum into the $[0,1]$ pixel range. In \emph{grayscale mode} we collapse the three decorrelated channels to their mean before the IRFFT and skip the final RGB-correlation matmul, producing a true $\mathrm{R}=\mathrm{G}=\mathrm{B}$ image. The Fourier spectrum is defined on a $64\times64$ grid and zero-padded to $256\times256$ before the IRFFT. Optimization runs for 3000 gradient steps per restart at this single resolution; ablations at higher resolutions and with three-stage curricula are reported in Appendix~\ref{app:ablations}.

\subsection{Loss}
With target ROI vertex set $\mathcal{R}$, all other cortical vertices $\bar{\mathcal{R}}$, predicted per-vertex activation $\hat{y}_v$, and Fourier spectrum $S$:
\begin{equation}
\mathcal{L} = -\frac{1}{|\mathcal{R}|}\sum_{v\in\mathcal{R}} \hat{y}_v \; + \; \beta \cdot \frac{1}{|\bar{\mathcal{R}}|}\sum_{v\in\bar{\mathcal{R}}} \hat{y}_v \; + \; \lambda_{\text{fft}} \cdot \|S\|^2
\end{equation}
With $\beta = 1.0$ and $\lambda_{\text{fft}} = 10^{-3}$, this is the \textbf{global lift loss}: maximize the target ROI mean activation minus the mean predicted activation elsewhere, with a mild spectral-energy regularizer to discourage pathological high-frequency content. We chose this objective because naive ROI maximization (the target term alone) tends to produce stimuli that drive many regions simultaneously rather than the target specifically: TRIBE v2's per-vertex predictions are correlated across nearby ROIs, so an unconstrained optimum exploits that shared variance. Subtracting mean activation over the rest of cortex penalizes solutions whose drive is generic, encouraging directions that are selective rather than merely large. The spectral-energy term is a mild regularizer against pathological high-frequency texture; $\lambda_{\text{fft}} = 10^{-3}$ was the smallest value at which we observed visible suppression of such textures without affecting low-frequency structure.

\subsection{Restarts and evaluation}
\label{sec:eval}
For each ROI we run 5 restarts with seeds $42 + 1000k$ for $k \in \{0,1,2,3,4\}$. We evaluate each restart by its \emph{final target activation} --- the mean of the encoder's predicted activation over the target ROI vertices at the last optimization step --- and treat the restart with the highest final target activation in each ROI column as the \emph{primary result} for the cross-ROI activation tables (Table~\ref{tab:selectivity}). The full restart grid is retained to inspect basin diversity across runs and is shown alongside the primary cells in Figure~\ref{fig:all-rois-grid}. As a baseline we also run the prediction pipeline on 5 random-noise images per ROI; the gap between the optimized stimulus's target activation and this 5-seed noise mean is reported as the \emph{lift}.

\subsection{Hardware}
All experiments were run on a single consumer RTX 3090 (24\,GB). Compute was the dominant bottleneck of this work: V-JEPA 2 ViT-G forward passes at fp16 saturate the card, restart counts were capped at 5 per ROI for budget reasons, and several natural extensions (full Glasser parcellation, video-axis optimization, larger restart ensembles) were deferred on compute grounds rather than scientific ones.

\section{Results}

The primary result is the cross-ROI grid of optimized stimuli (\S\ref{sec:cross-roi}): a visible V1$\to$V4 progression, and MT, FFA, and PPA each produce a distinctive signature. \S\ref{sec:quant} reports the corresponding cross-ROI activation matrix; \S\ref{sec:mt-frozen-motion} isolates the MT case, where static-only optimization still recovers implied-motion content; \S\ref{sec:superstim} quantifies the gap to a natural face photograph.

\subsection{Cross-ROI qualitative comparison}
\label{sec:cross-roi}

\begin{figure*}[tbp]
    \centering
    \includegraphics[width=0.95\linewidth]{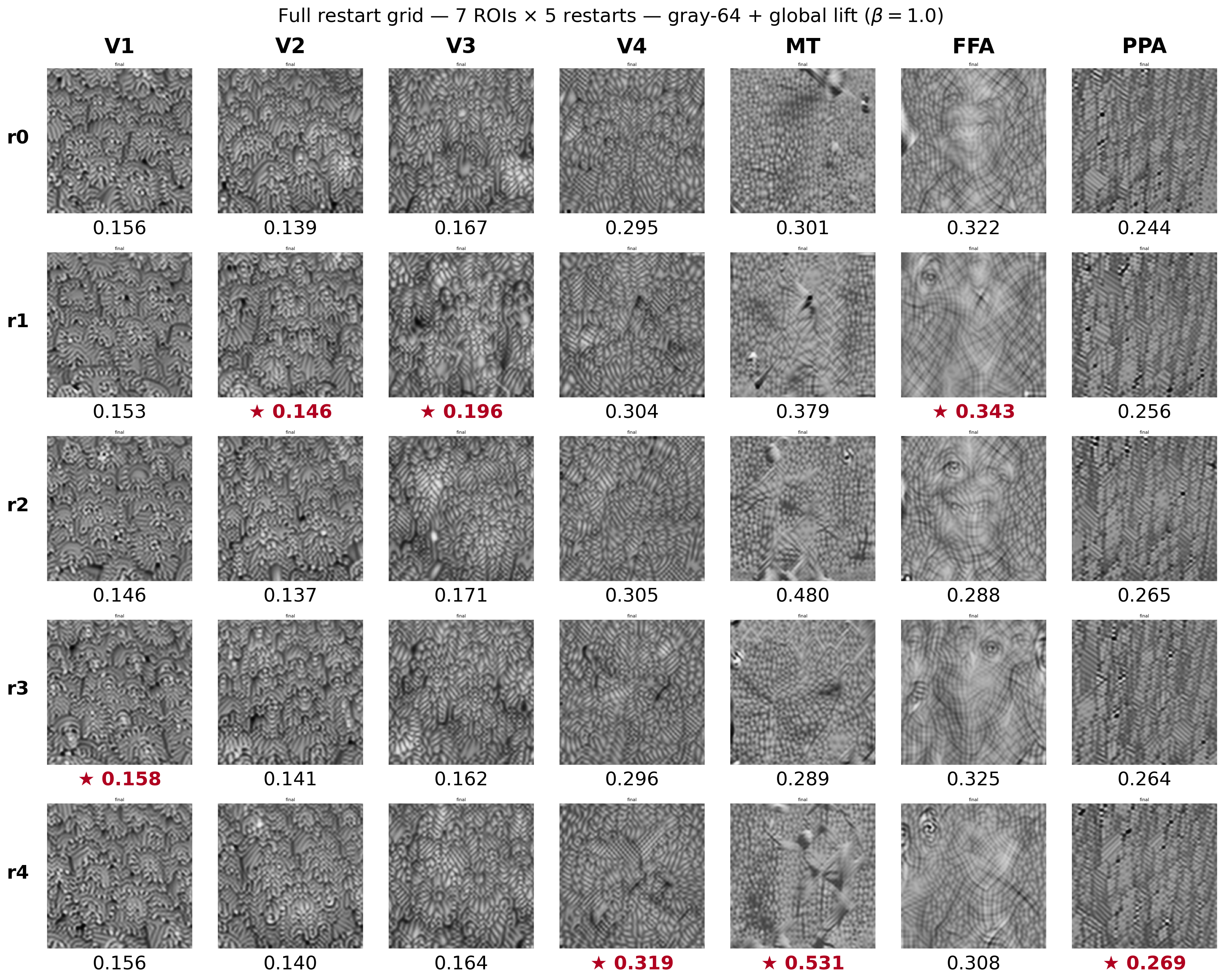}
    \caption{All 35 optimized stimuli: 7 target ROIs (columns) $\times$ 5 random-seed restarts (rows), under identical hyperparameters (gray-64 + global lift, $\beta=1.0$, $\lambda_{\text{fft}}=10^{-3}$, 3000 steps). The number printed below each panel is the predicted mean activation of the target ROI for that restart (TRIBE v2 z-scored units); the restart with the highest target activation in each column is marked with a star ($\star$). The FFA/r2 cell has been replaced with the clearest face the pipeline has produced (see \S\ref{sec:interp-gap}). Across V1$\to$V4 the optimized stimuli show a visible progression of increasing complexity and spatial scale; MT produces frozen-motion streaks; FFA produces face features; PPA produces consistent rectilinear lines. No region-specific hyperparameters were used --- all differences are driven by the choice of target ROI alone.}
    \label{fig:all-rois-grid}
\end{figure*}

Reading Figure~\ref{fig:all-rois-grid} column by column:

V1$\to$V4 shows a visible progression. The optimized stimuli become progressively larger in scale and more organized in structure across these four columns: V1 produces the densest fields of small-scale oriented edges and Gabor-like swirls; V2 is similar with slightly more contour junctions; V3 shows a noticeably larger scale; V4 is the most organized with mid-scale curves roughly $2$--$3\times$ the spatial scale of V1. V1 and V2 are the closest pair and individual restarts are not always distinguishable between them, but the gradient from V1 to V4 is apparent. The cross-activation structure of Table~\ref{tab:selectivity} confirms tuning overlap among V1--V3 alongside this progression.

All five restarts for MT produce sharp radial or diagonal streak patterns reminiscent of frozen optic flow, despite the pipeline optimizing \emph{static} single-frame inputs. MT also produces the largest target activation (0.49) and lift ($+0.70$) of any region (\S\ref{sec:mt-frozen-motion}).

FFA produces varied face-like content: eye-like regions, nose ridges, mouth/jaw outlines. Geometry differs between restarts but facial features appear in all.

PPA produces consistent rectilinear textures across all five restarts: parallel, predominantly horizontal and diagonal lines, with the tightest within-column consistency of any ROI. We refrain from a stronger interpretation: the stimuli look like line patterns rather than recognizably like scenes or architecture.

Every regime's qualitative signature is stable across all five random seeds, with variation restricted to specific geometry rather than category.

\subsection{Quantitative cross-ROI selectivity}
\label{sec:quant}

Table~\ref{tab:selectivity} reports target activation, lift over a random-noise baseline (mean across 5 noise seeds per ROI), and predicted activation across all seven candidate ROIs for each optimized stimulus.

\begin{table*}[!t]
\centering
\caption{Per-stimulus predicted activation across all seven candidate ROIs. \emph{Target act.} is the mean predicted activation over the target ROI vertices for the optimized stimulus (the ``final target activation'' of \S\ref{sec:eval}); \emph{Lift vs random} is that value minus the 5-seed random-noise baseline activation in the same ROI; remaining columns give predicted activation in each candidate ROI under the same stimulus. \textbf{Bold} = target ROI for that row. The target is the most-activated ROI in every row \emph{except} V3, whose stimulus drives V4 (0.251) more strongly than V3 itself (0.179) --- reflecting the tight V3/V4 coupling in V-JEPA~2's features and the strong cross-activation among early-visual ROIs more generally. Off-diagonal structure tracks known anatomical/functional relationships (\S\ref{sec:discussion-cross-roi}).}
\label{tab:selectivity}
\small
\begin{tabular}{lrr|rrrrrrr}
\toprule
Target & Target act. & Lift vs random & V1 & V2 & V3 & V4 & MT & FFA & PPA \\
\midrule
\textbf{V1}  & 0.155 & +0.179 & \textbf{0.155} & 0.123 & 0.086 & 0.087 & $-$0.448 & $-$0.102 & 0.022 \\
\textbf{V2}  & 0.138 & +0.143 & 0.158 & \textbf{0.138} & 0.119 & 0.126 & $-$0.320 & $-$0.029 & 0.023 \\
\textbf{V3}  & 0.179 & +0.183 & 0.123 & 0.127 & \textbf{0.179} & 0.251 & 0.000 & 0.123 & 0.032 \\
\textbf{V4}  & 0.292 & +0.248 & 0.059 & 0.053 & 0.128 & \textbf{0.292} & $-$0.015 & 0.144 & 0.073 \\
\textbf{MT}  & \textbf{0.490} & \textbf{+0.700} & $-$0.076 & $-$0.045 & 0.040 & 0.144 & \textbf{0.490} & 0.152 & 0.068 \\
\textbf{FFA} & 0.339 & +0.359 & $-$0.084 & $-$0.047 & 0.006 & 0.120 & 0.189 & \textbf{0.339} & $-$0.147 \\
\textbf{PPA} & 0.266 & +0.237 & 0.023 & 0.027 & 0.036 & 0.132 & $-$0.299 & $-$0.125 & \textbf{0.266} \\
\bottomrule
\end{tabular}
\end{table*}

TRIBE v2 predicts approximately z-scored blood-oxygen-level-dependent (BOLD) signal, so its outputs are relative to a zero mean: positive values indicate predicted activation above baseline, negative values indicate deactivation below it. PPA deactivates for faces, FFA for place-focused stimuli. The ``Lift vs random'' column uses 5 random-noise images as a proxy baseline; this proxy is not true cortical rest, since noise drives MT to $-0.21$ (no coherent motion) and FFA slightly negative ($-0.02$). A lift of $+0.36$ for FFA therefore corresponds to predicted activation of $+0.34$ above true baseline rather than above rest; this is of particular note for MT, where the noise baseline is strongly negative.

\begin{figure*}[!t]
    \centering
    \includegraphics[width=0.7\linewidth]{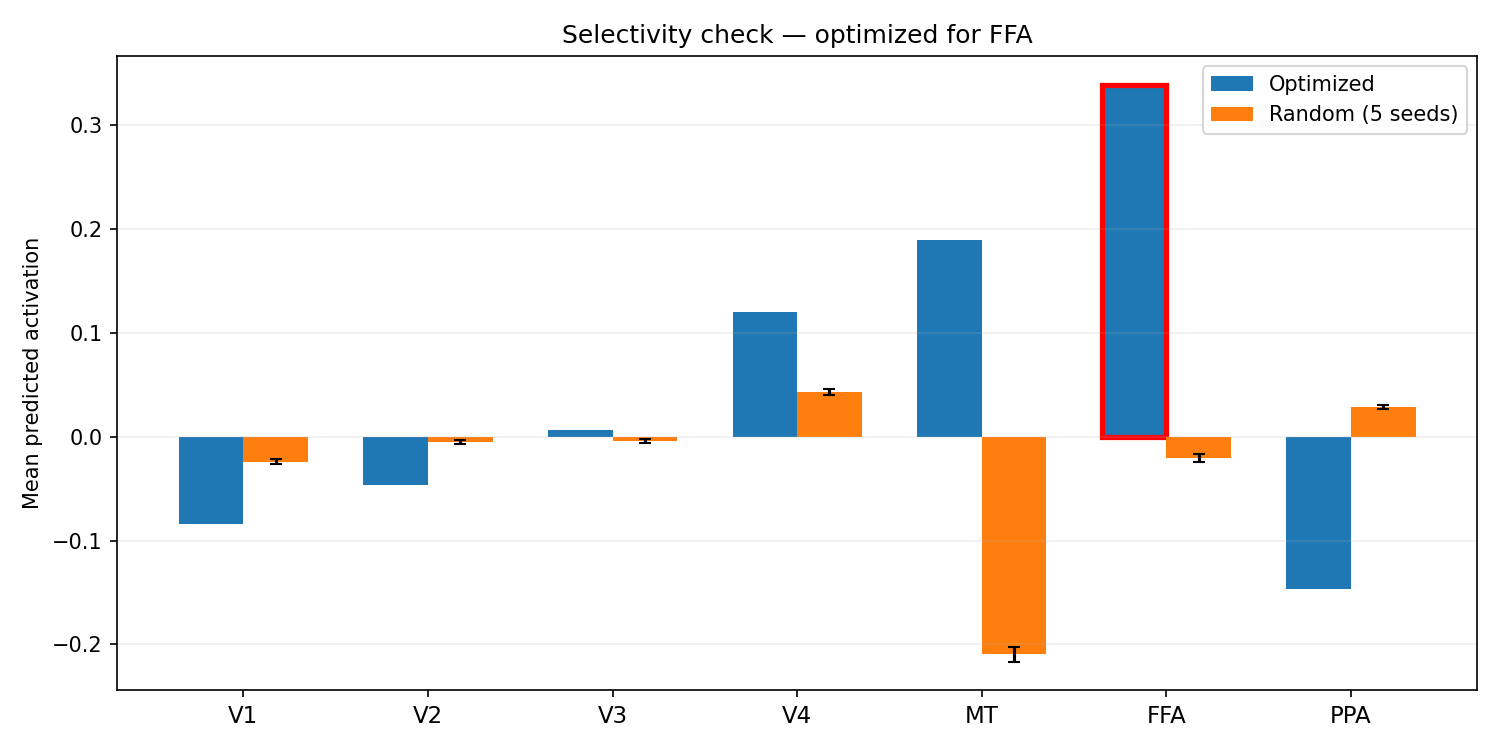}
    \caption{Per-ROI predicted activation for the FFA-optimized stimulus (blue) vs.\ the 5-seed random-noise baseline (orange). FFA (highlighted) is driven far above baseline; V4 and MT also increase moderately; PPA is suppressed well below baseline.}
    \label{fig:selectivity}
\end{figure*}

\subsection{MT produces ``frozen motion'' from static optimization}
\label{sec:mt-frozen-motion}

The most surprising result in Figure~\ref{fig:all-rois-grid} is the MT row. MT is canonically motion-selective --- its neurons are tuned to direction and speed of moving stimuli --- and our pipeline optimizes single-frame inputs tiled into a video, with no temporal degrees of freedom. \emph{A priori}, MT feature visualization under these constraints might have been expected to degenerate or produce a null result.

Instead, all five MT restarts produce sharp radial or diagonal streak patterns: elongated local orientations, converging lines, and bands that read as frozen optic flow or long-exposure photographs of moving scenes. MT also achieves the highest target activation (0.49) and lift ($+0.70$) of any region tested.

\subsection{Super-stimulus effect: optimized vs.\ natural face}
\label{sec:superstim}

\begin{figure*}[!t]
    \centering
    \begin{subfigure}[b]{0.22\linewidth}
        \centering
        \includegraphics[width=\linewidth]{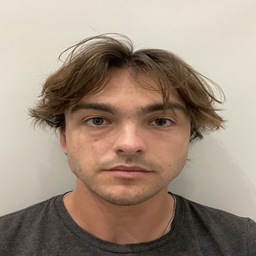}
        \caption{$+0.080$}
        \label{fig:real-face}
    \end{subfigure}\hspace{0.04\linewidth}
    \begin{subfigure}[b]{0.22\linewidth}
        \centering
        \includegraphics[width=\linewidth]{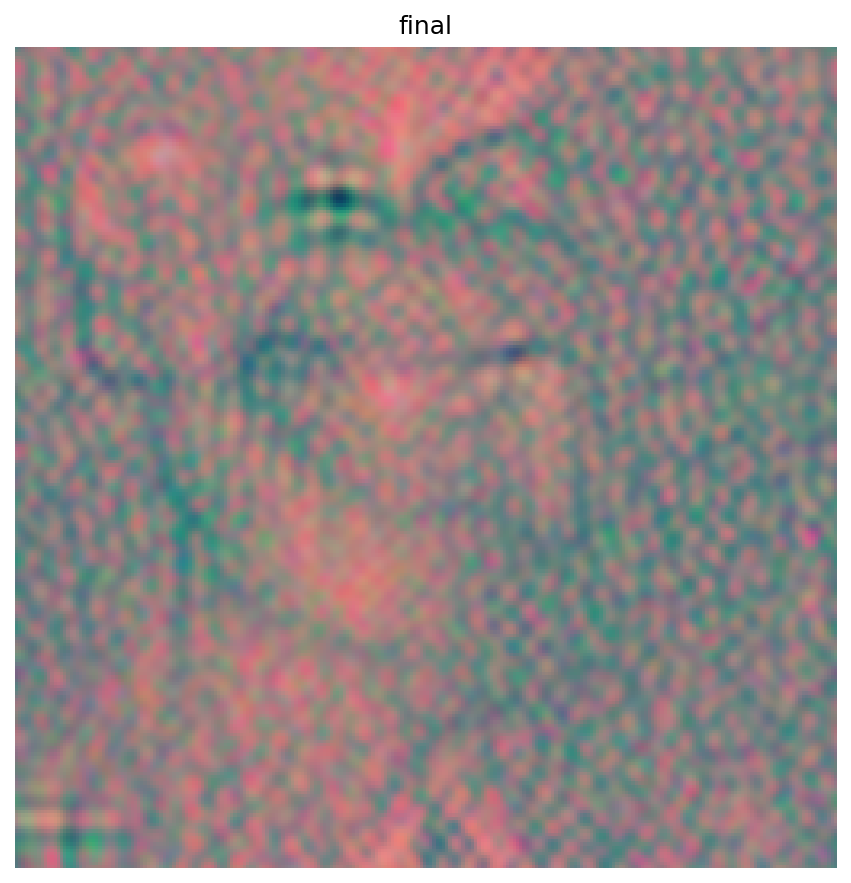}
        \caption{$+0.343$}
        \label{fig:ffa-color-breakdown}
    \end{subfigure}
    \caption{Natural face photograph (left, predicted FFA activation $+0.080$) versus the color-64 optimized FFA stimulus (right, predicted FFA activation $+0.343$). The natural face is a frontal portrait resized to $256\times256$ and tiled to 64 identical frames, used as the photograph row of Table~\ref{tab:superstim}. The optimized stimulus uses \emph{matched gray-64 methodology but with full colour} (single-stage $64\times64$ Fourier spectrum, 3000 steps, global-lift loss, $\lambda_{\text{fft}}=10^{-3}$). Colour expands the optimization parameter space by $3\times$ relative to grayscale, and the optimizer exploits the extra slack to inject high-frequency adversarial texture that drives FFA on top of the face-canonical drive --- visible as the pink/teal tint and high-frequency overlay. The face structure (eyes, nose, mouth) is still recognizable underneath, but the activation number includes an adversarial component the gray-64 result does not (\S\ref{sec:grayscale}).}
    \label{fig:face-and-color}
\end{figure*}

As a reference frame for our activations, we ran the prediction pipeline on two natural-face references: a stock vector line-drawing of a front-facing human face, and a colour portrait photograph. Both were resized to $256\times256$, converted to grayscale, and tiled to 64 identical frames to match the format of our optimized and noise stimuli; predictions were otherwise generated with the same frozen V-JEPA~2 + TRIBE v2 stack. Table~\ref{tab:superstim} reports the primary gray-64 number ($+0.339$); Figure~\ref{fig:face-and-color} displays the color-64 ablation variant ($+0.343$) alongside the natural-face reference, since the colour version is more visually interpretable as a face --- see \S\ref{sec:grayscale} for why both are reported.

\begin{table}[!t]
\centering
\caption{Optimized FFA stimuli drive the predicted region harder than the natural faces we tested, consistent with feature visualization producing adversarial super-stimuli rather than canonical exemplars. Natural-face comparisons are $n=1$ each and meant only to bracket the magnitude of the optimized drive.}
\label{tab:superstim}
\begin{tabular}{lr}
\toprule
Stimulus                       & FFA activation \\
\midrule
Random noise                   & $-0.020$       \\
Vector illustration of face    & $+0.039$       \\
Photograph of real face        & $+0.080$       \\
\textbf{Optimized FFA stimulus} & $\mathbf{+0.339}$ \\
\bottomrule
\end{tabular}
\end{table}

The optimized stimulus drives FFA roughly four times harder than the photograph (the classic feature-visualization super-stimulus phenomenon, in which gradient ascent finds patterns that exceed anything in the natural distribution \citep{szegedy2013intriguing,goodfellow2014explaining,walker2019inception,bashivan2019neural}). This does not mean the model is treating our stimulus as a face in any everyday sense; it means the optimizer has found a point in input space that hits V-JEPA 2's FFA-like tuning directions harder than photographs can.

\section{Ablations}\label{sec:ablations}

We ablate three components of the training recipe (Appendix~\ref{app:ablations}): a low-resolution-first curriculum that improves FFA selectivity ratio from $3.59\times$ to $13.33\times$ by committing the optimizer to global structure before high-frequency degrees of freedom become available; grayscale-only optimization, which gives a comparable raw FFA activation to the colour variant ($0.339$ vs.\ $0.343$) but without the high-frequency adversarial overlay visible in Figure~\ref{fig:face-and-color}b; and a comparison between global lift and targeted suppression of co-activated ROIs, where global lift is simpler and empirically stronger on primary metrics. We adopt gray-64 with global lift for the primary cross-ROI grid because its activation number is a more faithful measure of face-canonical drive.

\section{Discussion}
\label{sec:discussion}

\subsection{What the probe tells us about TRIBE v2}
TRIBE v2 was trained to minimize prediction error on naturalistic video fMRI, without any explicit supervision that FFA should respond to faces, MT to motion, or V1 to edges. Nonetheless, the probe recovers several tuning signatures cleanly at a qualitative level: V1$\to$V4 produces a visible progression of increasing spatial scale and feature complexity up the ventral stream, MT yields implied-motion streaks, and FFA yields face features. This is a stronger claim than prediction accuracy alone supports: the encoder has not merely fit the training fMRI time series, it has recovered the input$\to$region mapping well enough that running the pipeline in reverse reproduces the coarse tuning structure a neuroscientist would expect.

V1, V2, and V3 stimuli sit close together along the progression and cross-activate strongly in Table~\ref{tab:selectivity}. This is consistent with their known tuning overlap --- V2 and V3 inherit feature structure from V1 with incrementally larger receptive fields --- and the probe should be read as tracing a gradient rather than crisply partitioning these regions. The V3 row is diagnostic in the other direction: its stimulus drives V4 (0.251) more strongly than V3 itself (0.179), suggesting V-JEPA~2's features treat V3's tuning as a variant of V4's rather than as something independent.

We deliberately do not push the PPA interpretation beyond ``consistent rectilinear patterns.'' PPA stimuli look like parallel line fields; they do not obviously look like scenes or architecture. That the lines are predominantly straight and that PPA is canonically scene-selective is consistent but not in itself strong evidence for a category-level label. Of all seven ROIs, PPA is the one we are least comfortable interpreting --- a useful outcome given that one role of an interpretability probe is to expose where its target encoder is harder to read.

\subsection{Cross-ROI activation structure}
\label{sec:discussion-cross-roi}
Four structural features of Table~\ref{tab:selectivity} are worth highlighting. First, MT produces the largest target activation (0.490) and lift ($+0.700$) of any region; two factors combine, in that MT's baseline random-noise activation is strongly negative ($-0.21$) so the optimizer has more headroom to move, and MT contains only 38 vertices --- the smallest of our seven ROIs --- making its mean-activation objective easier to saturate than the larger ROIs. The visual character of the MT stimuli is more interesting than the raw number. Second, early-visual ROIs cross-activate strongly: the V2 stimulus drives V1 at 0.158, marginally stronger than the V1 stimulus drives itself (0.155), and the V3 stimulus drives V1 at 0.123 and V2 at 0.127. This is the numerical counterpart of the heavy V1--V3 tuning overlap that sits alongside the visual progression from V1 to V4. Third, FFA and PPA are cleanly separated from each other and from the rest: each suppresses the other ($-0.15$ for FFA$\to$PPA, $-0.13$ for PPA$\to$FFA), and the PPA stimulus pushes MT strongly negative ($-0.30$), consistent with PPA's preference for static scenes over motion. The FFA stimulus drives MT mildly positive ($+0.19$), suggesting V-JEPA~2 uses some motion-like features in its face-like drive. Fourth, V4 acts as a bridge between early visual and face-selective tiers: V4-optimized drives FFA at 0.144 (higher than it drives V1 or V2), and FFA-optimized drives V4 at 0.120, matching V4's classical role as a mid-level feature region feeding inferotemporal cortex.

\subsection{What the MT result suggests}
\label{sec:discussion-mt}
The MT finding is the cleanest demonstration of our central claim. MT is canonically motion-selective \citep{maunsell1983functional,born2005structure}, static images have no motion, and yet the pipeline recovers visual content that looks like frozen motion: radial streaks, converging lines, elongated orientations. The straightforward interpretation is that V-JEPA~2 was pretrained on video and has learned feature directions that fire on motion cues, including the \emph{implied} motion cues present in static images (streak patterns, oriented blurs, elongated local structures); TRIBE v2's MT readout has discovered these directions during fitting; and inverting the pipeline by gradient ascent recovers those implied-motion cues in pixel space. It is the same phenomenon that lets humans perceive motion in photographs of speeding cars or waterfalls --- a static-image effect that drives MT/MST in human fMRI studies as well \citep{kourtzi2000activation} --- the feature backbone has separated ``this image contains motion evidence'' from ``this image actually moves,'' and the MT readout is reading from the former. This is both a methodological success (a static pipeline recovered something meaningful for a motion region) and a limitation: we cannot distinguish true-motion tuning from implied-motion tuning without optimizing temporal content, which we have not done here. A video-based extension is the natural next step.

\subsection{Activation maximum $\neq$ interpretability maximum}
\label{sec:interp-gap}
Within the FFA column of Figure~\ref{fig:all-rois-grid}, final target activations vary across restarts, but visual interpretability varies more: the highest-activation restart is not unambiguously the most face-like. The cell occupying FFA/r2 is visually the cleanest face the pipeline has produced (clearly defined eye, symmetric brow ridge, nose, mouth), despite not being the column's activation maximum. CUDA fp16 non-determinism induces trajectory divergence even with identical seeds, but the functional category of the solution (face-like) is stable across runs. Activation magnitude is therefore a weak proxy for visual interpretability, and any downstream procedure that picks a ``canonical'' stimulus by argmax over restarts will sometimes pick a less recognizable one. We recommend selecting representative stimuli by inspection over a restart ensemble, not by raw activation.

\subsection{Limitations}
Several limitations constrain the interpretation of every result above. The optimized stimuli are extrema of the encoder's response surface (\S\ref{sec:superstim}): gradient ascent finds patterns that exceed anything in the natural distribution, and reasoning from these patterns to what a region ``really encodes'' risks overfitting to adversarial features of the composed pipeline. TRIBE v2 was trained on 64-frame clips and our tile-to-video mode is mildly out of distribution; the MT result (\S\ref{sec:mt-frozen-motion}) suggests this is workable for implied-motion content, but a video-axis extension would let us distinguish true-motion from implied-motion tuning. Finally, CUDA fp16 non-determinism limits pixel-level reproducibility, and $n=5$ restarts per ROI is small for formal statistics; we report restart variation but do not attempt significance testing on the lift-vs-noise metric for that reason.

\subsection{Future work}
The natural extensions are deferred on compute rather than scientific grounds: a sweep over the full Glasser parcellation to ask empirically how many functionally distinct tuning directions TRIBE v2 exposes; optimizing along the temporal axis to distinguish true-motion from implied-motion tuning in MT and to address motion-rich downstream regions; clustering optimized stimuli to discover groups of regions that share tuning; and presenting optimized stimuli to human observers or comparing them to held-out natural fMRI to make claims about cortex rather than the encoder. A larger restart ensemble would also support formal significance testing on lift vs.\ noise.

\section{Conclusion}
Held-out prediction accuracy is a necessary but insufficient evaluation of a brain encoder. Feature visualization --- gradient ascent on the encoder's predicted ROI activation --- is a complementary qualitative probe that asks whether the encoder has recovered the functional organization of cortex rather than merely fit fMRI time series. Applied to TRIBE v2 across seven cortical regions, identical hyperparameters produce a visible progression of increasing spatial scale and feature complexity across V1$\to$V4, plus three distinctive downstream regimes: implied-motion streaks for MT, face features for FFA, and consistent rectilinear line patterns for PPA. The MT result is particularly direct evidence of the method's reach: the encoder's motion readout can be inverted into recognizable motion-evoking imagery even when the stimulus format excludes literal motion. The probe is simple, differentiable, and applicable to any brain encoder built on a differentiable backbone --- the same test on future encoders will help distinguish those that have recovered neuroscience from those that merely predict it.

\bibliography{refs}
\bibliographystyle{icml2026}

\appendix
\section{Ablation details}
\label{app:ablations}

\subsection{Low-resolution-first curriculum}
\label{sec:lowres}
Progressing through resolutions $64 \to 128 \to 256$ during optimization --- rather than starting at full resolution --- meaningfully improves FFA selectivity: the selectivity ratio (target activation divided by mean activation across all other ROIs) rises from $3.59\times$ at full-resolution start to $13.33\times$ with the low-resolution-first curriculum. The intuition: the optimizer commits to global structure at low resolution before high-frequency degrees of freedom become available to ``cheat'' with pathological texture. We also experimented with the same three-stage curriculum at the primary grayscale setting and found it not obviously better than the single-stage $64\times64$ version on aggregate selectivity; we therefore report single-stage gray-64 as the primary configuration.

\subsection{Grayscale-only optimization avoids adversarial high-frequency exploits}
\label{sec:grayscale}
Running the full budget at $64\times64$ in pure grayscale with the global lift loss ($\beta=1.0$) gives FFA activation $0.339$ and lift $0.359$, with a simple single-stage training loop. The matched-methodology colour version (\emph{color-64}: same single-stage $64\times64$ Fourier spectrum, same step budget, same loss, just with the colour channels left independent) reaches $0.343$ --- a comparable raw number, but the way it gets there is different. Colour expands the optimization parameter space by $3\times$ (three Fourier channels rather than one), and the optimizer exploits the extra slack to inject high-frequency texture that drives FFA via adversarial directions in V-JEPA~2's feature space rather than purely face-canonical features. The result is the pink/teal-tinted, high-frequency-overlaid image of Figure~\ref{fig:face-and-color}b: face structure recognizable underneath, but adversarially polluted on top. This is the same phenomenon as the low-resolution-first lever in \S\ref{sec:lowres}: restricting the parameter space stops the optimizer from cheating its way to activation. We adopt gray-64 for the primary cross-ROI grid (Figure~\ref{fig:all-rois-grid}) not because the activation number is higher but because that number is a more faithful measure of face-canonical drive.

\subsection{Targeted vs.\ global suppression}
An earlier experiment used targeted suppression of V4 and MT (the two ROIs most co-activated by naive FFA maximization) with $\beta=0.3$. This gives slightly better suppression of V4 and MT specifically ($0.11$ and $0.05$ vs.\ $0.12$ and $0.19$ in the primary run) but slightly worse target activation ($0.289$ vs.\ $0.339$). Targeted suppression is more principled; global lift is simpler and empirically stronger on primary metrics. Both are defensible depending on what one wants to emphasize.

\end{document}